\documentclass[conference]{IEEEtran}
\IEEEoverridecommandlockouts
\usepackage{cite}
\usepackage{amsmath,amssymb,amsfonts}
\usepackage{algorithmic}
\usepackage{graphicx}
\usepackage{multirow}
\usepackage{makecell}
\usepackage{textcomp}
\usepackage{enumitem}
\usepackage{xcolor,colortbl}
\def\BibTeX{{\rm B\kern-.05em{\sc i\kern-.025em b}\kern-.08em
    T\kern-.1667em\lower.7ex\hbox{E}\kern-.125emX}}

\usepackage[all]{background}
\usepackage{stackengine}
\setstackEOL{\\}
\setstackgap{L}{\normalbaselineskip}
\SetBgContents{\color{gray}{\tiny \Longstack{PREPRINT - To appear at VLSID 2024}}}
\SetBgPosition{4.5cm,1cm}
\SetBgOpacity{1.0}
\SetBgAngle{0}
\SetBgScale{1.8}

\begin{document}
\bstctlcite{IEEEexample:BSTcontrol}

\definecolor{Gray}{gray}{0.85}
\definecolor{LightCyan}{rgb}{0.88,1,1}

\newcolumntype{a}{>{\columncolor{Gray}}c}
\newcolumntype{b}{>{\columncolor{LightCyan}}c}

\setlength\abovecaptionskip{0\baselineskip}
\setlength\belowcaptionskip{-1.4\baselineskip}
 
\def\@IEEEfigurecaptionsepspace{\vskip\abovecaptionskip\relax}%
\def\@IEEEtablecaptionsepspace{\vskip\abovecaptionskip\relax}%

\title{Experimental Validation of Memristor-Aided Logic Using 1T1R TaO$\rm _\textbf{x}$ RRAM Crossbar Array \vspace{-0.5cm}}

\author{Ankit Bende\IEEEauthorrefmark{1}\textsuperscript{\textsection}, \IEEEauthorblockN{Simranjeet Singh\IEEEauthorrefmark{1}\IEEEauthorrefmark{5}\textsuperscript{\textsection}, Chandan Kumar Jha\IEEEauthorrefmark{3}, Tim Kempen\IEEEauthorrefmark{6}, Felix Cüppers\IEEEauthorrefmark{6}, \\ Christopher Bengel\IEEEauthorrefmark{9}, André Zambanini\IEEEauthorrefmark{10}, Dennis Nielinger\IEEEauthorrefmark{10}, Sachin Patkar\IEEEauthorrefmark{5}, \\ Rolf Drechsler\IEEEauthorrefmark{3}\IEEEauthorrefmark{8},  
Rainer Waser\IEEEauthorrefmark{1}\IEEEauthorrefmark{6}, Farhad Merchant\IEEEauthorrefmark{2}, Vikas Rana\IEEEauthorrefmark{6} \vspace{1mm} \IEEEauthorblockA{\IEEEauthorrefmark{1}PGI-7, \IEEEauthorrefmark{6}PGI-10, \& \IEEEauthorrefmark{10}ZEA-2, Forschungszentrum Jülich GmbH, Germany, \IEEEauthorrefmark{9}IWE-2 RWTH Aachen, Germany,\\ \IEEEauthorrefmark{5}Indian Institute of Technology Bombay, India, \IEEEauthorrefmark{3}University of Bremen, Germany,\\ \IEEEauthorrefmark{8}DFKI GmbH, Germany,  \IEEEauthorrefmark{2}Newcastle University, UK} }
 \{a.bende, si.singh, t.kempen, a.zambanini, d.nielinger, v.rana, r.waser,\}@fz-juelich.de, \{simranjeet, patkar\}@ee.iitb.ac.in,  \\ {bengel@iwe.rwth-aachen.de}, \{chajha, drechsler\}@uni-bremen.de, farhad.merchant@newcastle.ac.uk \vspace{-0.6cm}}

\maketitle

\begingroup\renewcommand\thefootnote{\textsection}
\footnotetext{Equal contribution}
\endgroup

\begin{abstract}
Memristor-aided logic (MAGIC) design style holds a high promise for realizing digital logic-in-memory functionality. The ability to implement a specific gate in a MAGIC design style hinges on the SET-to-RESET threshold ratio. The TaO$\rm _x$ memristive devices exhibit distinct SET-to-RESET ratios, enabling the implementation of OR and NOT operations. 
As the adoption of the MAGIC design style gains momentum, it becomes crucial to understand the breakdown of energy consumption in the various phases of its operation. This paper presents experimental demonstrations of the OR and NOT gates on a 1T1R crossbar array. Additionally, it provides insights into the energy distribution for performing these operations at different stages. Through our experiments across different gates, we found that the energy consumption is dominated by initialization in the MAGIC design style. The energy split-up is 14.8\%, 85\%, and 0.2\% for execution, initialization, and read operations respectively.   

\end{abstract}

\section{Introduction}
Memristive devices, such as resistive random access memory (RRAM), offer a solution to the von Neumann bottleneck by implementing operations within memory itself~\cite{Sebastian2020nature,kempen}. One approach to implementing operations in memory is designing digital logic gates exploiting two distinct states - the high resistive state (HRS) and low resistive state (LRS) of an RRAM. These states are correspondingly mapped to logic ``0" and logic ``1" respectively. Depending on the input combination stored as a resistive state, the output memristor state can switch from one state to another, representing a logical output value. Several methods for achieving digital logic-in-memory (LiM) have been suggested in the literature. Various stateful and non-stateful logic techniques have been presented in the literature such as IMPLY~\cite{kvatinsky2013memristor}, FELIX~\cite{gupta2018felix}, majority logic~\cite{DDH+:2023},  and memristor-aided logic (MAGIC)~\cite{KBL+:2014}. Amongst all the techniques, MAGIC stands out as a popular choice because it stores the output in the form of the memristor's state itself, representing stateful logic.

Experimental validation of MAGIC gates has recently been achieved using fabricated valence change memory (VCM) devices~\cite{hoffer2020experimental,Hoffer:stateful}. However, this study specifically focuses on passive crossbar architectures, which suffer from sneak path currents and scalability challenges~\cite{ZIDAN:sneak-path}.  The passive crossbars also encounter difficulties during device formation, requiring significant initial current. To address these issues and enhance forming capabilities, a solution involves incorporating a transistor in series with the memristive device. This configuration creates a 1T1R cell, effectively mitigating sneak path problems. The 1T1R cell enables precise current control at the individual device level, offering enhanced control for the MAGIC operations. Despite the increase in physical footprint, the 1T1R configuration renders the overall system more scalable and allows for better control~\cite{ZIDAN:sneak-path}.

\begin{figure}[!t]
    \centering
    \includegraphics[width=0.9\linewidth]{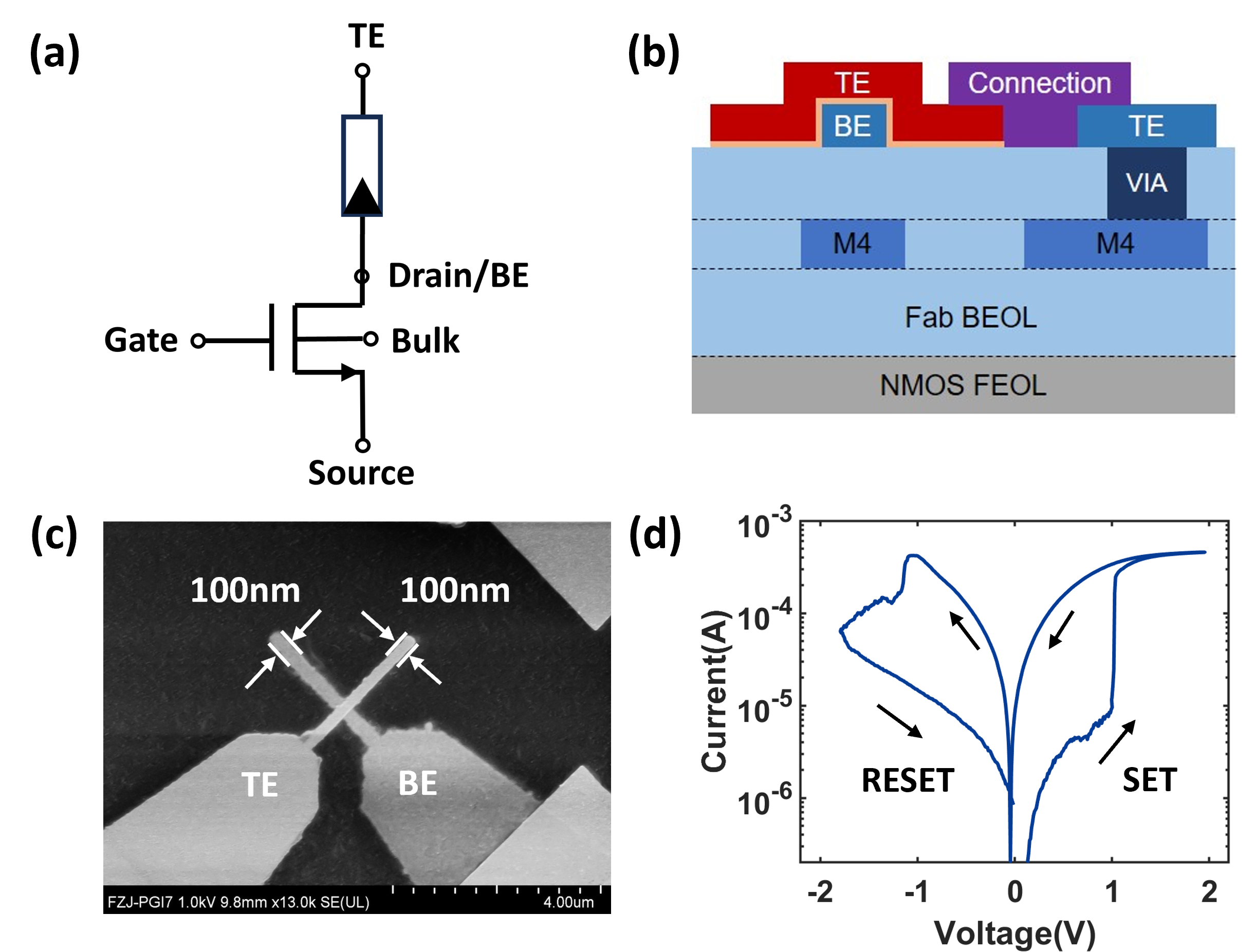}
    \caption{ (a) 1T1R cell schematic (b) schematic cross-section of CMOS integrated memristive device, (c) SEM image of fabricated memristive devices, and (d) I-V switching characteristics of a 1T1R cell after forming.\vspace{-5mm}}
    \label{fig:1T1R}
\end{figure}


The MAGIC design style offers the potential for a variety of logic gates. However, the availability of specific logic gates is contingent upon the SET and RESET switching thresholds, which are directly influenced by the material stack used in the fabrication process~\cite{Wald:design-methodology}. The RRAM device based on Pt/TaO$\rm _x$/W/Pt stack offers implementation of OR, NIMP, and 2-cycled XOR  gates and has been demonstrated using 1R passive devices~\cite{hoffer2020experimental}. Additionally, in this specific stack configuration, the output device is initialized to the HRS (logic ``0") state, and the input combination, along with execution voltage, determines its transition to the LRS (``1") from the HRS state.

To the best of our knowledge, for the first time, this paper demonstrates the implementation of MAGIC gates on a \emph{1T1R TaO$\rm _x$ RRAM crossbar array} and illustrates the realization of both OR and NOT gates, which can be effectively combined to implement any Boolean operation as they are functionally complete. Additionally, the paper offers a detailed breakdown of energy consumption during each operation phase within the MAGIC design style. Given the potential application of the MAGIC design style in creating general-purpose processing units, it becomes imperative to examine its design from an energy consumption perspective~\cite{EHA+:2021}. Such an analysis is crucial for gaining insights into its viability in future technologies. The following are the contributions of this paper:

\begin{itemize}
    \item Fabrication of TaO$\rm _x$ RRAM devices and their integration with CMOS to make the 1T1R crossbar array. 
    \item Experimental validation of OR and NOT gates based on MAGIC design style on fabricated 1T1R crossbar array.
    \item Comprehensive energy assessment and breakdown of energy consumption during the MAGIC operations.
\end{itemize}
The rest of the paper is organized as follows. We provide a background of the used technique in Section~\ref{backnrw}. Section~\ref{experimetns} discusses the experimental methods used for logic implementation and energy estimations. The results obtained from the study are explained in Section~\ref{discuss}. Finally, we conclude the paper in Section~\ref{conc}.

\begin{figure}[t]
    \centering
    \includegraphics[width=\linewidth]{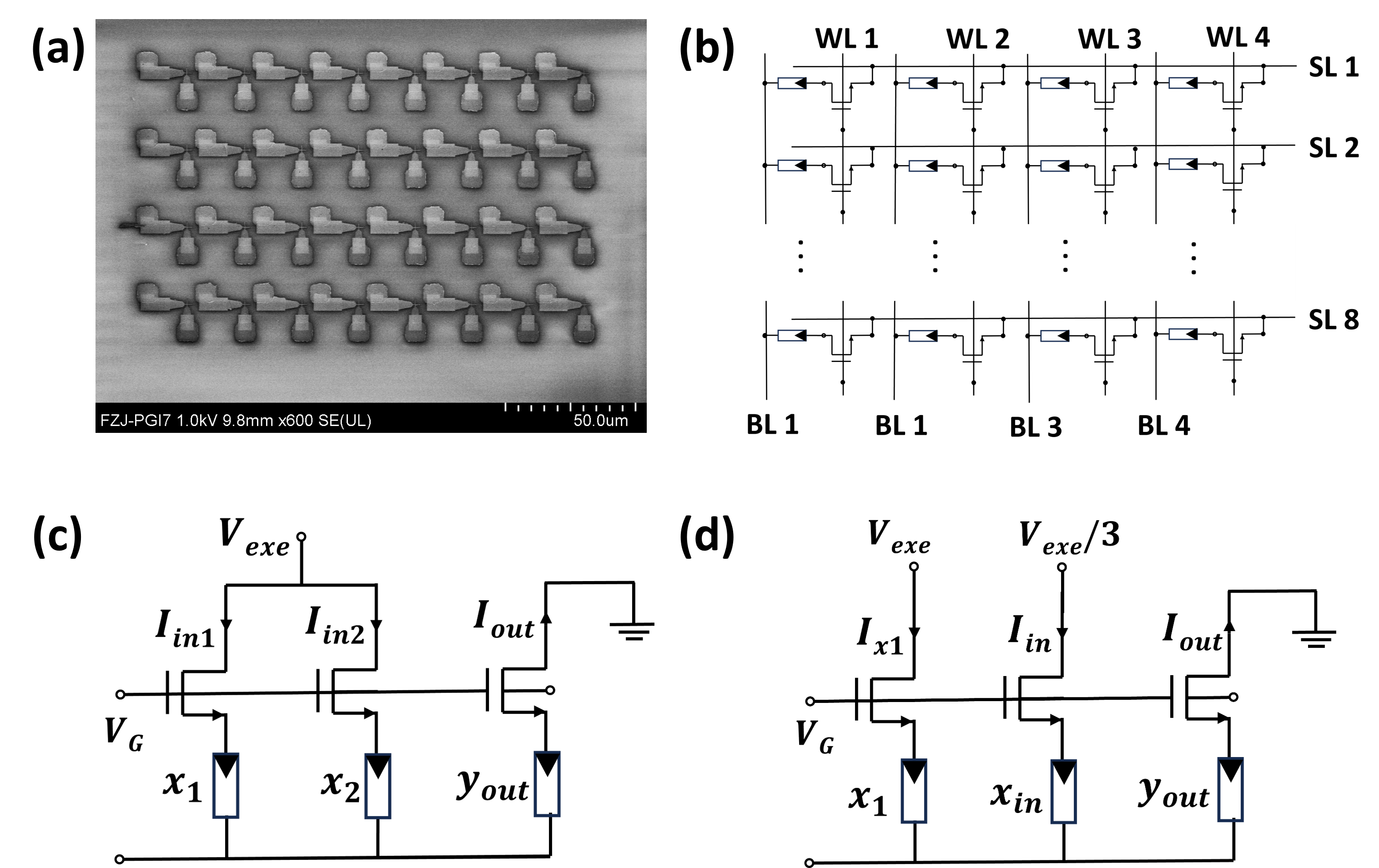}
    \caption{Logic gates mapping on fabricated 8x4 1T1R array. (a) SEM image of the fabricated 8x4 array (array orientation rotated by 90\textdegree), (b) Schematic of 8x4 1T1R array, (c) schematic of two input OR gate implemented in the array, (d) schematic of NOT gate implemented in the array.\vspace{-5mm}}
    \label{fig:gates}
\end{figure}


    

\section{Background and Related Work}
\label{backnrw}
\subsection{Memristive Devices}
Memristive devices have emerged as a significant advancement in non-volatile memory technology. Initially proposed as a concept by Professor Leon Chua in 1971~\cite{Chua:1971}, RRAM has gained prominence due to its unique ability to store data by modulating resistance states~\cite{Strukov:2008}. The resistance modulation is achieved by applying a  voltage across the terminals of the RRAM. In response, the resistance of the devices changes based on the magnitude and direction of the current flow.  Remarkably, the memristor preserves its resistance value even when devoid of power, safeguarding its data until a new voltage is applied, firmly establishing its status as a non-volatile memory element~\cite{Sebastian2020nature}.

These memristive devices can be interconnected to form a crossbar structure. However, when individual memristive devices are connected in a passive crossbar configuration, issues related to forming and sneak-path currents can arise. To mitigate these concerns, memristive devices are fabricated with a CMOS transistor in series, resulting in what is known as a 1T1R cell. In Fig.~\ref{fig:gates} (a), the SEM image of the fabricated 1T1R cell is shown, with multiple cells interconnected in an 8x4 (rows $\times$ columns) crossbar structure (fabrication detailed is discussed in Section~\ref{experimetns}). Fig.~\ref{fig:gates} (b) provides a schematic layout of the 1T1R crossbar array. The word lines (WLs) from 1 to 4 are linked to the gates of the devices connected in columns, while the source lines (SLs) from 1 to 8 are connected in a row-wise fashion, shorted to all the source pins of the transistors within the same row. The bit lines (BLs) from 1 to 4 are connected to the top electrode (TE) pin of each memristor within a column, as illustrated in Fig.~\ref{fig:gates} (b).

\subsection{MAGIC Design Style}
\label{MAGIC}

MAGIC represents a stateful logic methodology that employs crossbar-connected memristive devices to execute logic operations. Each memristor is programmable to two distinct states, HRS and LRS, which are subsequently mapped to logic ``0" and ``1" respectively. An initialization step is necessitated to facilitate the MAGIC operations, configuring the output memristor to its initial state. The range of achievable gates within these devices is contingent upon the SET-to-RESET switching threshold ratio. The SET-to-RESET voltage ratio for the device stack used in our study made logic OR, NIMP, and NOT gates attainable, diverging from the original NOR and NOT gates proposed in~\cite{KBL+:2014}. The output memristor ($y_{out}$) is always initialized to the HRS state rather than the LRS state. 

In the OR operation, an execution voltage ($V_{exe}$) is applied to the input memristors ($x_1$ and $x_2$). Simultaneously, the output memristor, initially set to the HRS state, is grounded, as depicted in Fig.~\ref{fig:gates} (c). Furthermore, the NOT operation necessitates three memristors but with different voltage values compared to the OR operation. In this context, one of the inputs is consistently initialized to LRS (designated as $x_1$ for clarity) in conjunction with the output memristor as shown in Fig.~\ref{fig:gates} (d). Additionally, distinct execution voltages ($V_{exe}$ and $V_{exe}/{3}$) are employed for the $x_1$ and $x_{in}$ memristors, respectively.

\subsection{Related Work}
The experimental validation of MAGIC design style on TaO$\rm _x$ RRAM devices using \emph{passive crossbars} has been demonstrated in existing literature~\cite{hoffer2020experimental}. Nonetheless, passive crossbars are plagued by the issue of sneak-path currents, which can disrupt the accurate reading of the final state. Additionally, forming processes in passive crossbars poses challenges. Non-stateful logic (e.g., scouting and majority logic) has been demonstrated using 1T1R  cells~\cite{Padberg2023experimental}. This paper represents the pioneering demonstration of \emph{stateful logic on a 1T1R TaO$\rm _x$ RRAM crossbar}.

Prior studies have indicated that energy consumption in the MAGIC design style is predominantly influenced by initialization energy~\cite{singh2023optimize}. However, it's essential to note that these findings are primarily derived from simulation studies. Furthermore, the presented energy figures are based on simulation models and pertain specifically to NOT and NOR gates based on the MAGIC design style. This paper takes strides towards calculating the energy consumption of OR and NOT gates in fabricated TaO$\rm _x$ RRAM devices.

\section {Experimental Methods}
\label{experimetns}
\subsection{Device Fabrication}
For the experimental validation of MAGIC gates, 1T1R-based active memristive arrays were fabricated and integrated with CMOS 180 nm technology provided by X-FAB. The dimensions of the memristive devices in the arrays are 100 $nm$ x 100 $nm$ and consist of Pt/TaO$\rm _x$/W/Pt stack. Fig.~\ref{fig:1T1R} (b) shows the schematic vertical cross-section of the fabricated device. Firstly,  the W plugs from the processed wafers were exposed, and the 25 nm thick Pt layer was deposited as the bottom electrode (BE) with DC sputtering. The BE layer was then patterned using electron beam lithography and back etching using reactive ion etching (RIE). A 7 nm thick TaO$\rm _x$ layer was then deposited by RF sputtering in Ar (77\%) and O$_2$ (23\%) gas mixture at 236W RF power followed by deposition of 13 nm thick W electrode using the DC sputtering. Subsequently, a 25 nm thick Pt layer was deposited as TE using the DC sputtering. Finally, the deposited switching oxide and TE stack were patterned using electron beam lithography and RIE-based back etching. Fig.~\ref{fig:1T1R} (c) shows the SEM image of the fabricated 1T1R TaO$\rm _x$ RRAM device. 

 
\subsection{Electrical Characterization Setup}

The electrical characterization of the 1T1R memristive devices was carried out using Keithley 4200 SCS. Fig.~\ref{fig:1T1R} (a) shows the schematic of the fabricated 1T1R memristive device with different terminals. The memristive device in its pristine state needs a one-time forming step, which involves the creation of a conductive filament in the switching oxide by applying a positive voltage across the memristor. The current through the memristor during the electroforming process is controlled by applying an appropriate DC gate voltage to avoid permanent breakdown of the oxide. To realize digital (logic in memory) LiM using these devices, four distinct operations are necessary for any logic operation, which are discussed below. 

\begin{itemize}[leftmargin=*]

    \item \textbf{SET Operation:} The SET operation involves changing the device's state from HRS to LRS. This is achieved by applying a positive ramp voltage from 0 to 1.8V at the TE electrode while maintaining a constant DC voltage of 1.6V on the gate terminal to limit the current flowing through the memristive to ~500$\mu A$. The drain and bulk pins of the transistor are connected to the ground.


    \item \textbf{RESET Operation:} The RESET operation switches the device from LRS to HRS by applying a positive ramp voltage from 0 to 2V at the source terminal, while keeping TE and bulk grounded. A minimum current requirement is crucial to dissolve the filament, and if not met, the device remains in LRS. Achieving this current requirement entails applying the maximum allowed voltage of 5V at the NMOS gate, opening the channel for current conduction, and affecting the minimum transistor size usable with memristors for SET and RESET processes.
    \item \textbf{Execution Operation:} During the execution operation, specific voltage configurations are applied to memristors for executing the OR and NOT operations. Once the input memristors are loaded with the input, the execution voltage applied across them decides the logic operation being executed on them. During this operation, the output memristor is connected to the ground. The detailed voltage value is discussed in Section~\ref{OR_explaination} and~\ref{NOT_explaination}.

    \item \textbf{Read Operation:} The read operation serves to determine the current state of the memristor. A read voltage of 0.5V is applied at the TE and the source and the bulk terminals are pinned to the ground. A gate voltage of 3.3V is applied at the gate to open the NMOS channel during read operation.
\end{itemize}

\begin{figure}[t]
    \centering
    \includegraphics[width=1\linewidth]{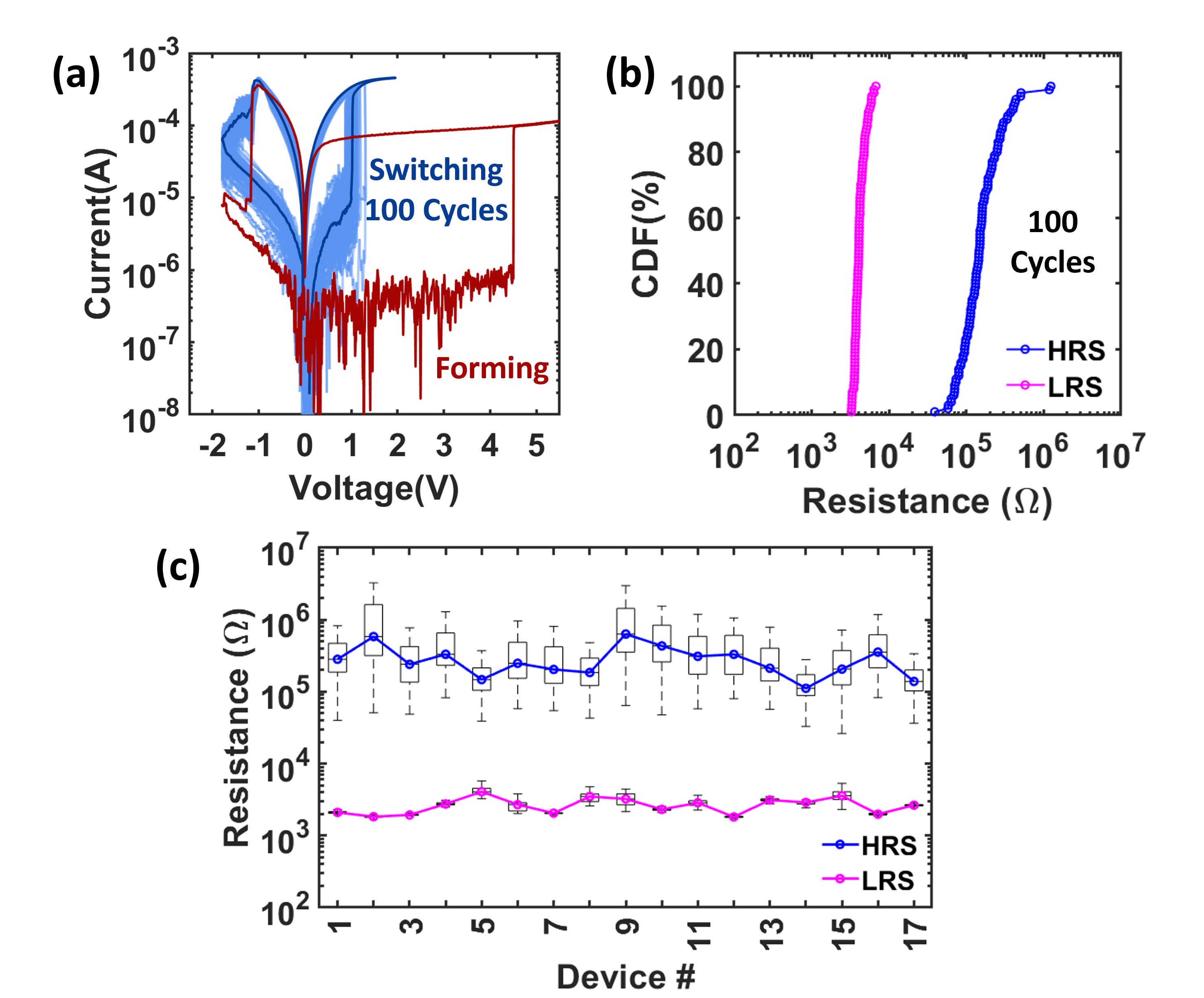}
    \caption{Electrical characterization of TaO$\rm _x$ RRAM devices. (a) Forming and median I-V curve for 100 switching cycles.  (b) Cycle-to-cycle variability for 100 switching cycles. (c) Device-to-device variability for 17 devices with 100 switching cycles each. \vspace{-5mm}}
    \label{fig:etest}
\end{figure}

Through skillful control of these operations, it becomes possible to design any arbitrary logic configuration on the crossbar. In this specific instance, these voltage sequences are combined to generate OR and NOT gates on the crossbar. However, these voltage patterns can also be harnessed to execute complex circuits sequentially or implement architectures resembling single instruction multiple data. Subsequently, we look into the outcomes achieved by sequentially applying these voltages to achieve the desired operations.



 

 

\section{Results and Discussion}
\label{discuss}

Within this section, we unveil the outcomes derived from our experimental examinations and analyses, offering valuable insights into the energy consumption associated with the implementation of MAGIC OR and NOT gates.

\subsection{1T1R TaO$\rm _x$ RRAM Switching Characteristics}
The initial forming and subsequent 100 switching cycles are shown in Fig.~\ref{fig:etest} (a). The memristive devices exhibit counterclockwise switching characteristics. Fig.~\ref{fig:etest} (b), shows the CDF plot of HRS and LRS states obtained by switching the 1T1R devices 100 times. The device exhibits low cycle-to-cycle (C2C) variation with a consistent HRS/LRS ratio of 10. For evaluating device-to-device (D2D) variability, around 17 devices were switched 100 times each as shown in Fig.~\ref{fig:etest}~(c).

\begin{figure}[t]
    \centering
    \includegraphics[width=\linewidth]{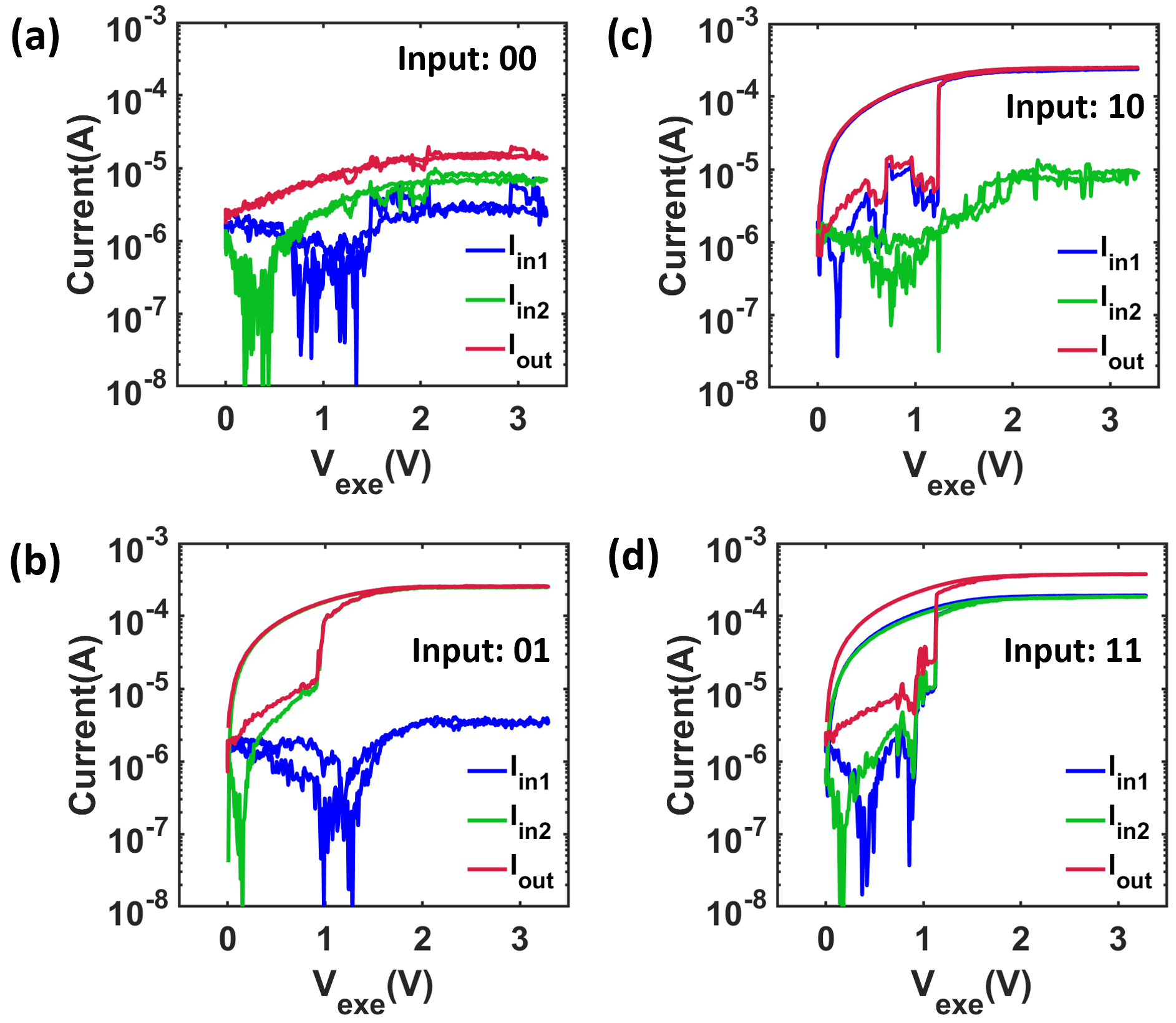}
    \caption{Execution operation in OR gate for (a) input ``00", (b) input ``01", (c) input ``10" and (d) input ``11". For input ``00" there is no sharp change in I$_{out}$ implying no change of y$_{out}$ state. For the rest of the input combinations, y$_{out}$ undergoes the SET process by drawing the majority of the current through the input memristors that are in the LRS state.    \vspace{-5mm}}
    \label{fig:or_exe}
\end{figure}

\begin{figure}[t]
    \centering
    \includegraphics[width=0.85\linewidth]{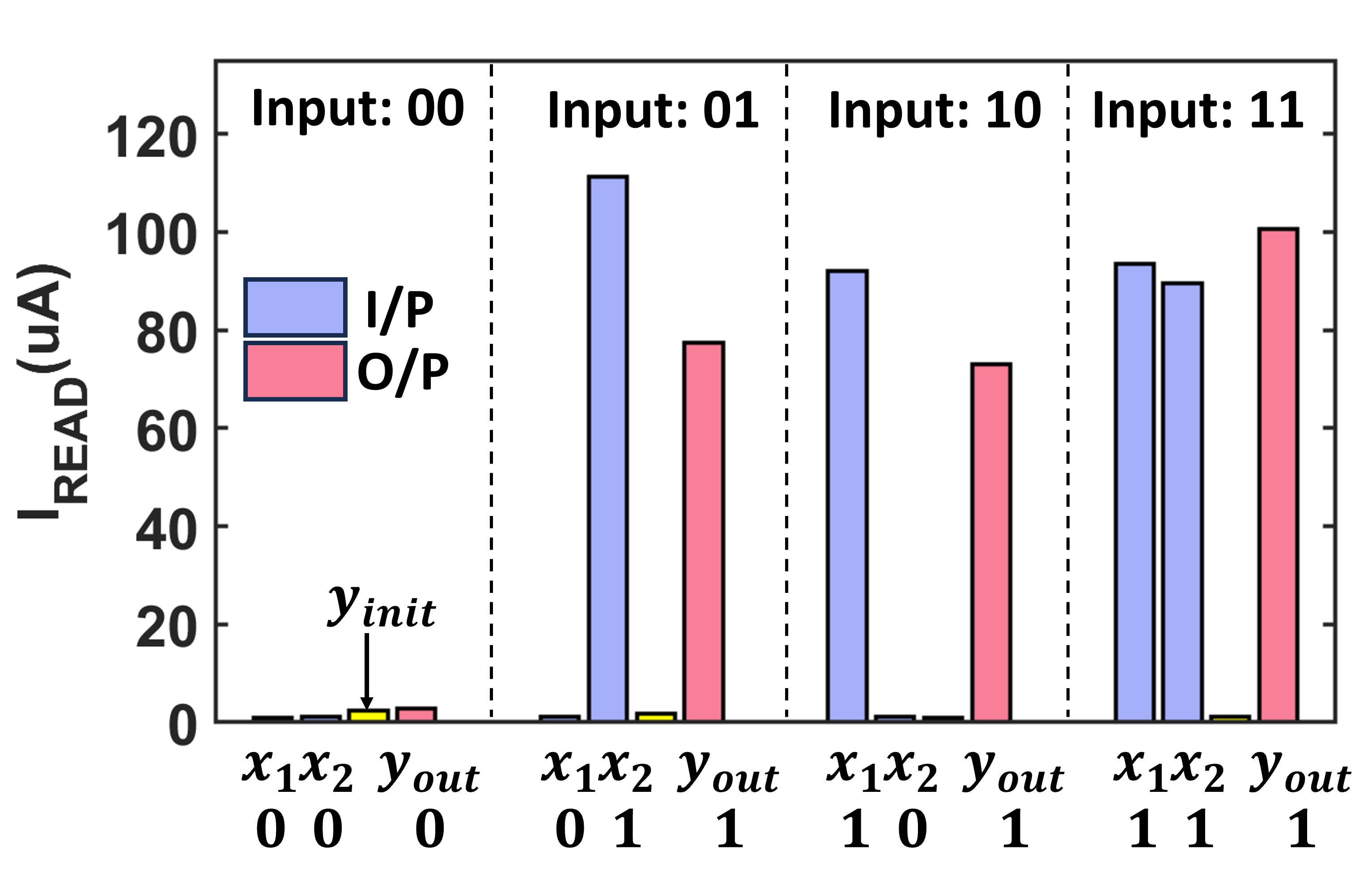}
    \caption{Logic OR operation. Output read currents for the different combinations of input read currents.\vspace{-5mm}}
    \label{fig:or_iread}
\end{figure}

From the C2C and D2D variability plots, it is evident that the HRS state exhibits higher variability with resistances distributed over two orders of magnitude ranging from around 100K$\Omega$ to 1M$\Omega$. The high variability in HRS can be attributed to the stochastic or uncontrolled breaking of filament during the RESET process~\cite{hrs_var}. Although the devices exhibit variability in the HRS state, a \emph{consistent HRS/LRS ratio of 10} is obtained across all the tested devices and suffices for the implementation of logic gates. Next, we will discuss the execution of logic OR and NOT operations on these devices.
\begin{figure}[t]
    \centering
    \includegraphics[width=\linewidth]{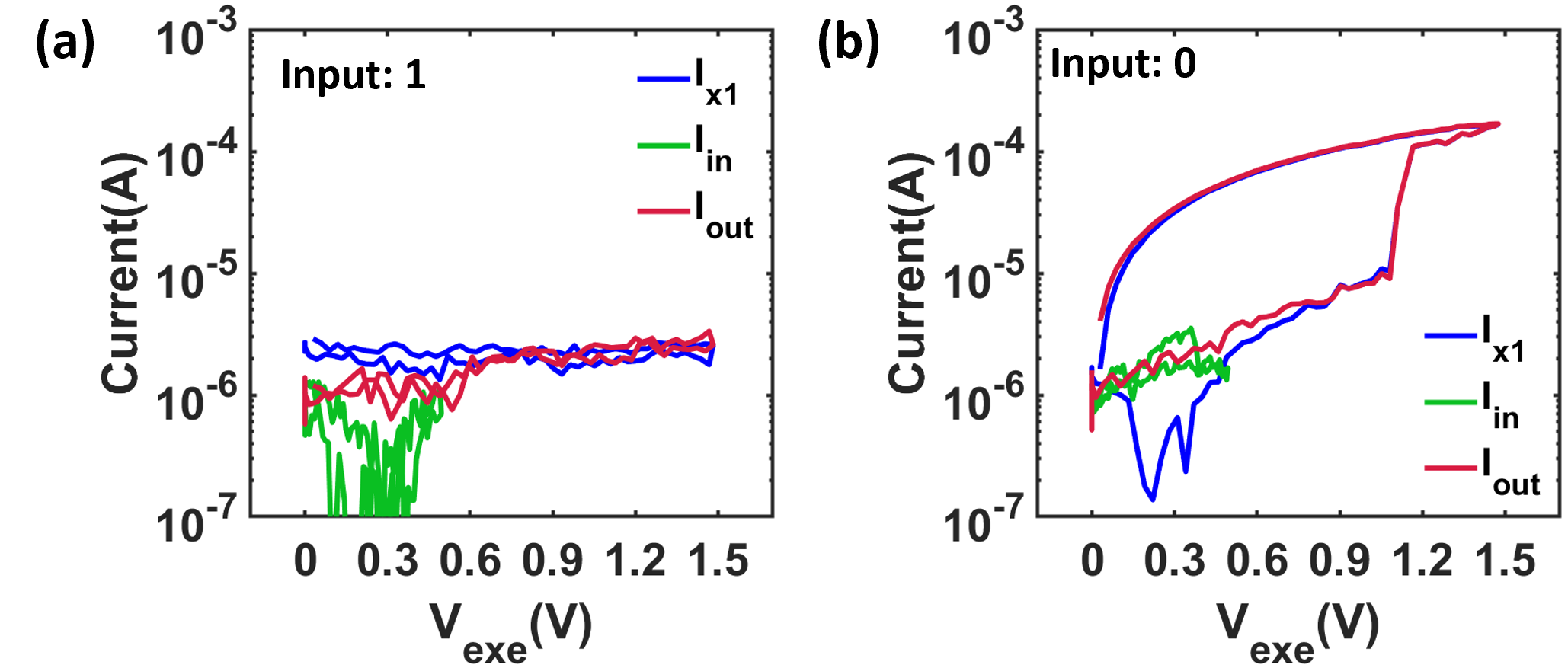}
    \caption{Execution operation in NOT gate for (a) input ``1", (b) input "0".\vspace{-10mm}}
    \label{fig:not_exe}
\end{figure}


\begin{table}[b]
    \centering
    \caption{Truth table for OR and NOT logic gate operation}
    \begin{tabular}{|a|a|c|b|c|a|c|b|}
    \hline
    \multicolumn{4}{|c|}{OR} & \multicolumn{4}{c|}{NOT} \\
    \hline
        \textbf{$x_1$} & $x_2$ & $y_{init}$ & $y_{out}$  & $x_1$ & $x_{in}$ & $y_{init}$ & $y_{out}$ \\
        \hline
            0    &  0 &  0 &  0    & -& -& - &- \\
        \hline
             0    &  1 &  0 &  1   & -& -& - & - \\
        \hline
             1    &  0 &  0 &  1   &  1    &  0 &  0 &  1    \\   
        \hline
             1    &  1 &  0 &   1   &  1    &  1 &  0 &  0    \\
        \hline         

    \end{tabular}
    \label{tab:or_tt}
\end{table}

\subsection{MAGIC OR Implementation}
\label{OR_explaination}
Fig.~\ref{fig:gates} (c) shows the schematic to implement the logic OR gate on the crossbar. To implement OR gate only three memristors are required so this needs to be mapped to the fabricated crossbar with a size of 8x4 as shown in Fig.~\ref{fig:gates} (c). To map the OR gate on the crossbar, three memristive devices sharing a common WL and a BL in the array are used to implement the gates. Firstly, the inputs of the OR gate are stored as the resistance state in $x_1$ and $x_2$. Subsequently, the output memristor ($y_{out}$) is initialized to the ``0" state.
Next, an execution voltage ($V_{exe}$) sweep from 0 to 3.3V at the TE terminal of $x_1$ and $x_2$ and the current at $y_{out}$ is monitored. During this cycle, the BL shared by three $x_1$, $x_2$ and $y_{out}$  is kept floating. While WL/$V_G$ is connected to a DC voltage of 3.3V and the SL of the output memristor is grounded. All the other unused WLs, SLs, and BLs are kept floating. Post-execution cycle, the state of $y_{out}$ is obtained by performing a READ operation. The memristor output currents during execution cycles for different inputs are shown in Fig.~\ref{fig:or_exe}. 

The truth table for the OR gate with the input and output states of the memristor are shown in Table~\ref{tab:or_tt}. In Table~\ref{tab:or_tt},  $y_{init}$ column shows the initialization state of the output memristor before the execution cycle. It can be inferred from the truth table that for successful OR gate operation, the $y_{out}$ changes its state for all combinations of inputs except for the input ``00" during the execution cycle. Fig.~\ref{fig:or_exe} shows the execution cycles for different input combinations. For the input ``00" case, both $x_1$ and $x_2$ are in HRS state represented by the current value $I_{in1}$ and $I_{in1}$, respectively. When an execution voltage is applied at the inputs, the current through the $y_{out}$ is limited by the parallel combination of $x_1$ and $x_2$. This current is not sufficient to drive the $y_{out}$ to the SET state. This is evident from the fact that no sharp change in output current ($I_{out}$) w.r.t applied $V_{exe}$ is observed in the $y_{out}$ as shown in Fig.~\ref{fig:or_exe}~(a). 

On the other hand, for the input combination of ``01", ``10", and "11", either one or both of the input memristors are in LRS states. This allows a sharp increase in $I_{out}$ during the execution cycle, contributing to the change of state of $y_{out}$ as shown in Fig.~\ref{fig:or_exe} (b), Fig.~\ref{fig:or_exe} (c) and Fig.~\ref{fig:or_exe} (d), respectively. In Fig.~\ref{fig:or_iread}, the read currents of output memristors for different combinations of input memristors are displayed, showing a successful logic OR operation.




\begin{figure}[t]
    \centering
    \includegraphics[width=0.7\linewidth]{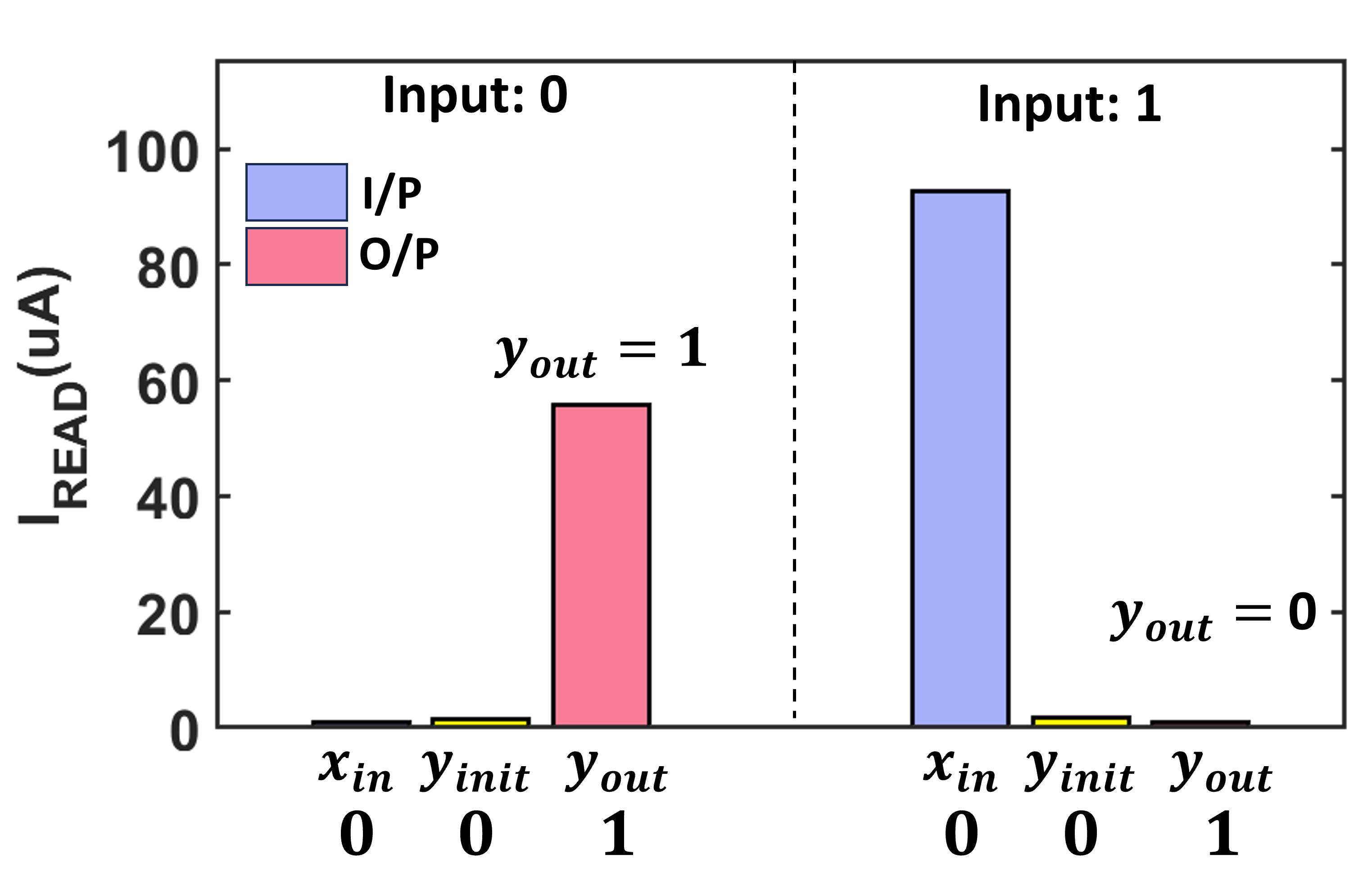}
    \caption{ Logic NOT operation. Output read currents for the different combinations of input read currents. \vspace{-5mm}}
    \label{fig:not_gate_iread}
\end{figure}

\begin{table}[b]
    \centering
    \caption{Number of SET and RESET operations required during initialization for OR gate implementation}
    \begin{tabular}{|c|c|c|}
    \hline
         & \multicolumn{2}{c|}{No. of initialization steps}  \\
        \hline
        Input & RESET & SET \\
        \hline
            00    &    3    &    0    \\
         \hline
         01 & 2 & 1 \\
         \hline
         10 & 2 & 1 \\
         \hline
         11 & 1 & 2 \\
         \hline
    \end{tabular}
    \label{tab:number_of_init}
\end{table}

\subsection{MAGIC NOT Implementation}
\label{NOT_explaination}
The schematic of the MAGIC NOT gate is shown in Fig.~\ref{fig:gates} (d). For performing NOT logic operation, memristor $x_1$ and $y_{out}$ are initialized to LRS and HRS state, respectively. Input memristor $x_{in}$ is initialized in accordance with the input. An execution ramp voltage of $V_{exe}$ from 0 to 1.5V is applied at the source terminal of $x_1$ whereas 1/3 $V_{exe}$ voltage is ramped at the $x_{in}$. Table~\ref{tab:or_tt} shows the truth table for NOT gate with different input combinations. For the successful operation of NOT gate, $y_{out}$ memristor should change its state when input ``0" is applied and should remain in ``0" state for input ``1". The currents in the output memristor during the execution cycle for different inputs are shown in Fig.~\ref{fig:not_exe}. 

The input voltage applied at $x_{in}$ is 1/3rd of the voltage applied at $x_1$. Because of this, during the input ``0", the potential difference across the $y_{out}$  memristor (currently in HRS state) is higher as compared to the $x_{in}$. This higher potential across $y_{out}$ allows it to undergo a transition from RESET to SET through $x_1$ (currently in LRS) state. This can be seen as a sharp rise in output current as depicted in Fig.~\ref{fig:not_exe} (b). On the other hand, for the case of input bit ``1", the $y_{out}$ memristor path becomes higher in resistance as compared to $x_{in}$ path and does not receive enough current to drive it into SET from RESET state as shown in Fig.~\ref{fig:not_exe} (a). This allows $y_{out}$ to change its state only when $x_{in}$ is in the HRS state. Fig.~\ref{fig:not_gate_iread} summarizes the output read currents for different input read current combinations and successfully demonstrates NOT gate.

\begin{table}[b]
    \centering
    \caption{Energy consumption in SET, RESET, and READ operations for full voltage ramp cycle}
    \begin{tabular}{|c|c|c|c|}
    \hline
        Operation & Voltage(sweep)  & Pulse duration ($\mu$ s) & Energy(nJ)  \\
        \hline
         SET & 0 to 2 V & 4 ms & 1300  \\
         \hline
         RESET & 0 to 1.8 V  & 3.6 ms & 312  \\
         \hline
         LRS Read & 0 to 0.3 V & 0.6$\mu$ s & 5.4 \\
         \hline
         HRS Read & 0 to 0.3V & 0.6$\mu$ s & 0.056  \\
         \hline
    \end{tabular}
    \label{tab:energy_init}
\end{table}

\begin{figure*}[t]
    \centering
    \includegraphics[width=0.9\linewidth]{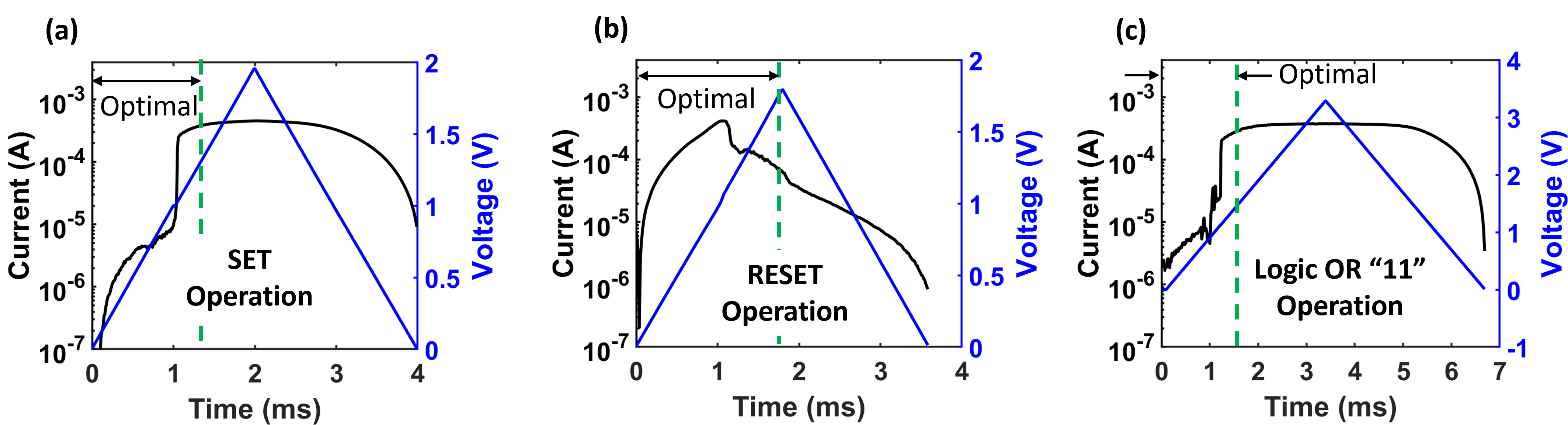}
    \caption{I,V-t curves for calculating the energy consumption of (a) SET process (b) RESET process (c) Logic OR "11" input execution.\vspace{-3mm}  }
    \label{fig:setresetor}
\end{figure*}

\subsection{Energy Calculations}
The energy consumption of logic operations is heavily dependent on the initialization as well as execution energies. In earlier works, researchers have typically calculated the energy consumption of a log-in memory system through a coarse-grained approach by multiplying the average energy of operations by the number of operations~\cite{pastmethod}. However, this method has been found to underestimate the actual energy consumption as it ignores the initialization energy involved during the operations~\cite{singh2023optimize,singh2023memspice}. However, these results are in the simulation and with different memristor models. Therefore, in the current study, a similar approach has been considered for real devices, for calculating the energy consumption of logic operations by taking execution as well as initialization energy into account.



\begin{table*}[t]
    \centering
     \setlength{\tabcolsep}{3.5pt}
       \caption{ Comparison of initialization energy to the total energy consumed during different input combinations for logic OR operations.} 
    \begin{tabular}{|c|c|c|c|c|c||c|c|c|c|c|}
      \hline
         OR& \multicolumn{5}{c||}{ Full Voltage Ramp cycle} & \multicolumn{5}{c|}{\textbf{Optimal}} \\
         \cline{2-11}
         \cline{7-9}
 Input&  Initialization & Execution & Read & \% Initialization &\% Read & Initialization & Execution & Read &\% Initialization &\% Read \\ 
  &  Energy (nJ) & Energy (nJ) & Energy (nJ) & Energy &Energy & Energy (nJ) & Energy (nJ) & Energy (nJ) & Energy &Energy \\
\hline
00 & 3900 & 139 & 0.1&97 &  0.002& \textbf{696} & \textbf{8} & 0.07& 99 & 0.003\\ \hline
01 & 2912 & 2455 & 5.4&54 &  0.1& \textbf{738} & \textbf{108} & 2.8& 87 & 0.1\\ \hline
10 & 2912 & 2300 & 5.4&56 &  0.1& \textbf{738} & \textbf{73} & 2.8&91 &  0.1\\ \hline
11 & 1924 & 3531 & 10.8&35 &  0.2& \textbf{780} & \textbf{134} & 5.6&85 &  0.2\\
\hline
    \end{tabular}
    \label{tab:energy}
\end{table*}

During OR operation, the different initialization steps involve numerous SET and RESET operations as mentioned in Table~\ref{tab:energy_init}. In the study, triangular wave sweep voltages are used to perform, SET, RESET, and READ as well as execution operations. The energy is obtained by multiplying the voltage waveforms with the sensed current and then integrating the product over the measurement time as per $\int_{0}^{t} v(t) \times  i(t) \,dt$, where $t$ is the pulse time considered for energy calculations.
The I,V-t curves corresponding to median SET and RESET operations are shown in Fig.~\ref{fig:setresetor}. Two approaches have been used for calculating the energy: a) using full voltage ramp cycle time and b) using optimum times. In the first technique, energy is calculated for the whole triangular voltage sweep. In contrast, in the case of optimum energy calculation, the SET, RESET, and execution times have been derived from the experimental data, and energy is calculated for the same as shown in Fig.~\ref{fig:setresetor}. 

The former technique gives actual energy consumption numbers but is an overestimate. Therefore, for more realistic energy values, the latter technique is used. Table~\ref{tab:energy} summarises the energy consumption for OR and NOT gate operation through both techniques. It is quite evident that for both the optimal and full voltage ramp cycle-based energy calculations, depending on the inputs of the OR operation, around 35-85\% of the energy consumption is constituted by the initialization energy, execution energy constitutes around 64.8-14.8\% while read energy only constitutes around 0.2\% of energy  Evidently, as the accuracy of energy calculation increases, the gap between total energy and execution energy also increases. Therefore, for accurate calculation of energy consumption for logic operations using the MAGIC technique, the initialization energy must be taken into consideration.


\section{Conclusion}
\label{conc}
In this paper, TaO$\rm _x$ RRAM devices were fabricated and integrated on a 180 nm CMOS substrate to create a 1T1R crossbar array. The fabricated devices consistently exhibited an HRS/LRS ratio of approximately 10, making them eligible for the implementation of MAGIC gates. Subsequently, we demonstrated the implementation of logic OR and NOT gates, and along with energy consumption values. Energy consumptions for logic OR and NOT operations were calculated by evaluating both the initialization and execution energies. It was found that the initialization energy played a significant contributing role in the overall energy consumption during logic implementation, similar to the trends observed in the simulation study.

\section*{Acknowledgments}
This work was supported in part by the Federal Ministry of Education and Research (BMBF, Germany) in the project NEUROTEC II under Project 16ME0398K, Project 16ME0399, German Research Foundation (DFG) within the Project PLiM (DR 287/35-1, DR 287/35-2) and through Dr. Suhas Pai Donation Fund at IIT~Bombay. Special thanks go out to Dr. Michael Schiek for his management work in the NEUROTECH II project.

\bibliographystyle{IEEEtran}
\bibliography{Bib}

\end{document}